\documentstyle[aps,prb]{revtex}
\input{epsf}
\begin{document}
\draft

\title{Statistical properties of the time evolution of complex systems. I}
\author{{\large P. Leb{\oe}uf and G. Iacomelli}} 
\address{Division de Physique Th\'eorique\footnote{Unit\'e de recherche des
Universit\'es de Paris XI et Paris VI associ\'ee au CNRS.}, Institut de
Physique Nucl\'eaire, 91406 Orsay Cedex, France\\} 
\date{\today}
\maketitle

\begin{abstract} 
The time evolution of a bounded quantum system is considered in the  framework
of the orthogonal, unitary  and symplectic circular ensembles of random matrix 
theory. For an $N$ dimensional Hilbert space we prove that in the large $N$
limit the return amplitude to the initial state and the transition amplitude to
any other state of Hilbert space are Gaussian distributed. We further  compute
the exact first and second moments of the distributions. The return and 
transition probabilities turn out to be non self-averaging quantities with a 
Poisson distribution. Departures from this universal behaviour are also  
discussed. 
\end{abstract}


\vspace{2cm}

\section{introduction}

During the last decades much theoretical progress has been made concerning the
physical properties of complex quantum systems. In the 1950s and 1960s,
following the pioneering work of Wigner, a statistical theory was developed
\cite{porter} to interpret the basic features of highly excited nuclear states.
Known today as the random matrix theory (RMT), it has had an impressive success
in  the description of a large class of systems, ranging from nuclei and atoms
down to  microwave chaotic cavities or transport properties of electrons in
metallic or  ballistic disordered quantum dots (for a review of the
developments see for instance
Refs.\onlinecite{bohigas,casati,leshouches2}). 

Most past studies focus on the spectral statistics of complex systems based on
linear wave dynamics. Our purpose here is to concentrate on the time  evolution
of such systems. We are particularly interested in specific quantum (or wave)
manifestations in the dynamics. If at $t=0$ the system is prepared in a given
state $|\psi(0)\rangle$, a quantity which has been extensively studied 
is the average of the return probability $|C(t)|^2 = | \langle 
\psi(0)|\psi(t)\rangle |^2$ (sometimes called the survival probability). This 
function,  which is accessible from experimental measurements as the modulus
squared of the Fourier transform of the spectrum, was studied by Leviandier et
al \cite{leviandier} in their analysis of spectral correlations in molecular
physics. Their basic finding is that the spectral rigidity is manifested as a
"hole" in the average of $|C(t)|^2$ for times shorter than the Heisenberg time
(proportional to the inverse of the mean level spacing), this test being more
robust than other ways of testing the spectral correlations. The technique was
also applied to the nuclear data ensemble \cite{lombardi} and the spectra
of superconducting microwave cavities \cite{alt}. From the theoretical point of
view, a rigorous interpretation of this phenomenon in the context of RMT was
given in Refs.\onlinecite{guhr,alhassid} (see also \onlinecite{gruver}). The
correlation hole was also obtained using supersymmetry techniques and
interpreted as a quantum dynamical echo in the time evolution of a wave packet
within a disordered mesoscopic sample \cite{prigodin}, and in acoustics
\cite{weaver}. In Ref.\onlinecite{prigodin} a direct dynamical experimental test 
of the correlation hole based on the transient currents through a disordered 
dot was also proposed, but such a direct test is still lacking.

Our purpose here is to describe the statistical properties of the return
probability as well as those of more fundamental quantities such as the return
amplitude $C(t)$ and the transition  amplitude to a different state
$|\chi\rangle$, $T(t) =\langle \chi|\psi(t)\rangle$. Our basic assumption is
that the system can be described, from a statistical point of view, by a random
matrix belonging to one of the three well known symmetry  classes of RMT. In
this  respect, and due to the Hilbert space rotational invariance of these
ensembles, the  theory applies to the relaxation process of a closed system for
which the rate of  exploration of phase space is sufficiently high in  order to
avoid localization effects. We show in the following that in the large-$N$
limit the real and imaginary part of $C(t)$ and $T(t)$ are Gaussian distributed
for any time. It then follows that their modulus squared (the return and
transition probabilities) are Poisson distributed and that moreover these are
non self-averaging quantities (in the sense that their fluctuations are of the
same order of magnitude as their mean value in the large-$N$ limit). This 
provides incidentally a proof of the results of Ref.\onlinecite{prange} where
the Poisson distribution was conjectured. Finally we obtain an exact
expression, valid for any $N$, for the form factor and thus for the average
return probability. This generalizes the asymptotic results obtained in
Refs.\onlinecite{guhr,alhassid,prigodin}. An intuitive interpretation of the 
results obtained in terms of a correlated random walk process in the complex 
plane is also given.

Deviations from the universal behaviour associated to a RMT description are
expected to occur  in particular for short times. These effects, which are not
initially included in our description, are briefly discussed in the last part
of the paper. Our results are moreover obtained for the circular ensembles of 
RMT. In the context of dynamical systems these ensembles are closely related
to the quantized version of classically chaotic maps \cite{maps}. In fact, the 
time-dependent Schr\"odinger equation of these systems is precisely given by 
Eq.(\ref{1}) below. We expect, however, that in the large-$N$ limit the results 
reported here will be general, applicable to any fully chaotic system. This is 
so because the asymptotic behaviour of Circular and Gaussian ensembles 
coincide. As we will discuss here and in more detail elsewhere \cite{il}, 
some differences may however remain, in particular in connexion with the 
discrete nature of time in the Circular ensembles.

Following Dyson \cite{dyson}, we start by representing a system not by its
Hamiltonian but by an $N\times N$ unitary matrix $U$ which determines the time
evolution of the system according to the equation 
\begin{equation}\label{1}
|\psi(n)\rangle = U^n |\psi(0)\rangle
\end{equation}
where $n=0,1,2,\ldots$ represents the time in units of an arbitrary interval of 
time which we fix to unity. The discrete nature of the time
evolution encoded in this equation is not essential for the moment, and 
the time  will be
rescaled to a continuous variable later on. The eigenvalues of $U$ are $N$
complex numbers lying on the unit circle 
\begin{equation}\label{2}
U |\varphi_k\rangle = \exp(i \theta_k) |\varphi_k\rangle \;\;\;\;
\;\;\;\;\;\;\;\; k=1,\cdots,N \ .
\end{equation}
Our main observables of interest are the amplitude of return 
\begin{equation}\label{3}
C(n)\equiv\langle\psi(0)|\psi(n)\rangle=\sum_{k=1}^N |a_k|^2 {\rm 
e}^{i n 
\theta_k}
\end{equation}
and the transition amplitude to a different state $|\chi\rangle$
\begin{equation}\label{4}
T(n)\equiv\langle\chi|\psi(n)\rangle=\sum_{k=1}^N a_k \bar{b}_k 
{\rm e}^{i n 
\theta_k} \ ,
\end{equation}
with $a_k=\langle\varphi_k|\psi(0)\rangle$ and
$b_k=\langle\varphi_k|\chi\rangle$. Due to the discrete and finite nature of
the spectrum these are quasiperiodic functions of time. However in
the large $N$ limit they will be  highly oscillatory functions, and our purpose
now is to study their statistical  properties.  A statistical analysis of these
amplitudes may be introduced by considering   $|\psi(0)\rangle$ as a  random
vector. However, this is not the way we proceed. More  relevant to experiments
is to fix the initial state vector and to consider the matrix $U$ as  a random
matrix belonging to one of the circular ensembles of random matrices
\cite{dyson,mehta}: (a) symmetric and unitary for systems having time reversal
invariance and  rotational symmetry or having time reversal invariance and
integral spin ($\beta=1$);  (b) unitary if there is no time reversal symmetry
($\beta=2$); (c) self-dual unitary  quaternion for half-integral spin systems
with time reversal invariance  ($\beta=4$). $C(n)$ and $T(n)$ thus become
random functions of the random variables $a_k,\ b_k$  and $\theta_k$ whose
distribution we wish to study. The state $|\chi\rangle$ will be considered to 
be statistically independent with respect to $|\psi (0)\rangle$, and thus the 
$b_k$'s independent from the $a_k$'s. For $\beta=4$ each eigenvalue in 
Eqs.(\ref{3}) and (\ref{4}) is in fact doubly degenerate \cite{dyson} (Kramers
 degeneracy) and $|a_k|^2$ represents in this case the sum of the square of the
amplitudes in the degenerate subspace $|a_k^{(1)}|^2 +|a_k^{(2)}|^2$ (and
analogously $a_k \bar{b}_k$ represents the sum $a_k^{(1)}  \bar{b}_k^{(1)} +
a_k^{(2)} \bar{b}_k^{(2)}$). 

A basic property of circular (as well as Gaussian) ensembles of random matrices
is their Hilbert space rotational invariance. There are two  basic
consequences of this invariance. The first one is the statistical independence
of the  eigenvalues with respect to the components of $|\psi (0)\rangle$: 
$ P(\vec{a},\vec{\theta})=P_1 (\vec{a})  \, P_2 (\vec{\theta})$, where
$\vec{a}=(a_1,\cdots,a_N)$, $\vec{\theta}=(\theta_1,\cdots,\theta_N)$  and $P$
represents the joint probability density for the corresponding ensemble. The
second consequence is the isotropic form of $P_1$ 
\begin{equation}\label{5}
P_1(\vec{a}) = \frac{2}{S_{\beta N}} \delta \left( \sum_{k=1}^N 
|a_k|^2 - 1\right) \ .
\end{equation}
The normalization constant $S_{\beta N}$ is the surface of a $(\beta N-1)$
dimensional sphere. The coefficients $a_k$ are real for $\beta=1$  and complex
for $\beta=2$ and $\beta=4$. We will not need the exact expression for $P_2
(\vec{\theta})$ but simply note that $P_2$ is invariant under a global phase
shift of the $\theta_k$. Thus the distribution of $C(n)$ depends only on its
modulus. 

\section{The distribution of the return and transition amplitudes}

We now show that the probability distribution of the real part of $C(n)=C_1 + 
i \ C_2$ is, in  the large $N$ limit, Gaussian. Similar steps can be
repeated for the imaginary part. From Eq.(\ref{3}) the distribution is defined 
as 
$$
{\cal P}(C_1)=< \delta\left[ C_1-\sum_{k=1}^N |a_k|^2 \cos (n \theta_k) \right] >
$$
where the brackets denote the ensemble average over the amplitudes and phases.
Using the integral representation of the delta function, the average over the 
amplitudes reduces to
$$
\prod_{k=1}^N \int d^2 a_k P_1 (a_k) \exp \left[- i \xi |a_k|^2 
\cos(n\theta_k) \right]
$$
where $P_1 (a_k)$ is the distribution of a single component. Because 
asymptotically the latter distribution is Gaussian, then 
$$
{\cal P}(C_1)\approx \ <\int\frac{d \xi}{2 \pi} {\rm e}^{i \xi C_1} 
\exp{\left[-\frac{1}{2} \sum_k\ln
\left(1+\frac{2 \ i \ \xi \cos (n \theta_k)}{N}\right)\right]}>
$$
where the remaining ensemble average is over the phases. In the large $N$ limit
we can expand the logarithm and keep only the $1/N$ term. The latter
equation then reduces to
$$
{\cal P}(C_1)\approx \ <\int\frac{d \xi}{2 \pi} 
\exp{\left[i \ \xi \left( C_1 - \frac{1}{N} \sum_{k=1}^N \cos (n \theta_k) 
\right) \right]}> \ ,
$$
i.e. the distribution of the function ${\tilde C}_1 (n) =\sum_k \cos (
n  \theta_k)/N$. This is a function that depends only on the phases (the 
powers of the traces of $U$). The distribution of the powers of the traces of 
unitary operators was discussed in Ref.\onlinecite{haake}, where it was shown 
to be Gaussian for low powers. More generally, it has been shown 
\cite{politzer} that  asymptotically the distribution of any "linear" function 
of the form $\sum_k f(\theta_k)$ is  Gaussian. Since ${\tilde C}_1 (n)$ is 
precisely of that form, this then proves that the real and imaginary part of 
the return amplitude  are Gaussian distributed. 

For the transition amplitude the proof is slightly different. For 
simplicity we limit here to the case $\beta=1$ (real coefficients); the other 
symmetries can be treated in the same way. The distribution of the real part of 
$T (n)=T_1 + i \, T_2$ is, by definition (cf Eq.(\ref{4}))
$$
{\cal P}(T_1)=< \delta\left[ T_1-\sum_{k=1}^N a_k b_k \cos (n \theta_k) 
\right] > \ .
$$
The use of the integral representation for the delta function and the 
computation of the integrals over the $a$'s and $b$'s now leads to
$$
{\cal P}(T_1)\approx \ <\int\frac{d \xi}{2 \pi} {\rm e}^{i \xi T_1} 
\exp{\left[-\frac{1}{2} \sum_k\ln
\left(1+\frac{\xi^2 \cos^2 (n \theta_k)}{N^2}\right)\right]}>
$$
where again the remaining average is over the phases. Keeping the leading order 
term in $1/N$ in the expansion of the logarithm it follows that
$$
{\cal P}(T_1)\approx \ <\int\frac{d \xi}{2 \pi} 
\exp{\left[i \ \xi T_1  - \frac{\xi^2}{2 N^2} \sum_{k=1}^N \cos^2 (n \theta_k) 
\right]}> \ .
$$
Finally, computing the integral over $\xi$ we arrive to
$$
{\cal P}(T_1)\approx \ < \frac{1}{\sqrt{2 \pi \sigma_{T_1}^2 (\vec{\theta})}}
\exp \left(-\frac{T^2}{2   \sigma_{T_1}^2 (\vec{\theta})} \right) > \ ,
$$
where
$$
 \sigma_{T_1}^2 (\vec{\theta}) = \frac{1}{N^2} \sum_{k=1}^N \cos^2 (n \theta_k)
\ .
$$
Because in the large $N$ limit the variance $ \sigma_{T_1}^2 (\vec{\theta})$ 
can be approximated by 
$$
 \sigma_{T_1}^2 \approx \frac{1}{2 \pi N} \int_0 ^{2 \pi} \cos^2 (n \theta_k) 
\, d  \theta = \frac{1}{2 N} (1+\delta_{n,0})  \ ,
$$
it then follows that the distribution of $T_1$ is Gaussian with a variance 
$ \sigma_{T_1}^2 $. Likewise, for the imaginary part of $T$ we also get a 
Gaussian distribution with a variance
$$
 \sigma_{T_2}^2 \approx \frac{1}{2 \pi N} \int_0 ^{2 \pi} \sin^2 (n \theta_k) d 
\ \theta = \frac{1}{2 N} (1-\delta_{n,0})  \ .
$$
We thus recover, as it should, Eq.(\ref{9}) below for the variance of $T$
obtained from a direct computation of the average return probability.

\section{The return and transition probabilities}

\subsection{Their average value}

Having established that the asymptotic distribution of the real and imaginary 
part of the return and transition amplitudes are Gaussian distributed,
we now consider their first and second moment of those distributions. 
Of particular interest is the second moment of the return amplitude. But before
that, let's consider the first moments. 

Since the ensemble average of $|a_k|^2$ is given by $<|a_k|^2>=\int |a_k|^2 P_1
(\vec{a}) d\vec{a} = 1/N$, then the average of the return
amplitude Eq.(\ref{3}) is 
$$
<C(n)>=\frac{1}{N} \int_{-\pi}^\pi d\theta \, {\rm e}^{i n 
\theta} R_1 (\theta)
= \delta_{n,0} \ , 
$$
where $R_1 (\theta)=N/2\pi$ is the level density. Therefore on average the
vector $|\psi(n)\rangle$ decorrelates immediately from the initial  state. This
is obviously due to the randomization of the phase of $|\psi(n)\rangle$. For
the  transition amplitude we find $<T(n)>=0, \ \forall n$. 

The second moment of the distribution of the return amplitude is the 
probability of return 
\begin{equation}\label{6}
|C(n)|^2 = \sum_{k=1}^N |a_k|^4 + \sum_{j\neq k}^N |a_j|^2 
|a_k|^2 {\rm e}^{i n 
(\theta_j - \theta_k)} \ .
\end{equation} 
As mentioned before, this quantity was considered in earlier work by different
authors. For the circular ensembles of RMT we obtain
\begin{equation}\label{7}
<|C(n)|^2> =\frac{N\delta_{n,0} + (\beta+2)/\beta- b_\beta^{(N)}(n)}
{(N+2/\beta)} \ ,
\end{equation} 
which is an exact expression. The function $b_\beta^{(N)}(n)=\frac{4\pi}{N} 
\int_0^\pi d\theta \cos(n\theta) T_2(\theta)$ is the Fourier transform of the  
two-level cluster function, the so-called spectral form factor \cite{mehta}. 
This function, which for $n=0$ is equal to one while for large times tends to 
zero, is responsible for the "correlation hole" mentioned in the introduction.
In fact, aside from the delta function at the origin, the numerator in 
Eq.(\ref{7}) goes from $2/\beta$ for short times up to its large-time 
stationary value $(\beta+2)/\beta$. Due to the periodicity of the circular
ensembles and the discreteness of time, this result differs from the one
obtained for the Gaussian ensembles (no convolution present, cf
Refs.\onlinecite{alhassid,il}). 

Previous works have considered the large $N$ limit of the average return
probability by employing the well-known large-$N$ behaviour of the form factor.
In order to have an expression valid for any dimension, we have computed the
form factor for any $N$, with the result 
\begin{eqnarray} \label{8}
b_1^{(N)}(\tau) &=& 2 \left(1-\tau-\frac{\tau}{2}\sum_{p=1}^{N-n} 
\frac{1}{(N+1)/2-p} \right) \Theta(1-\tau) - \frac{2}{N} \sum_{p=1/2}^{(N-1)/2} 
\left(\frac{p^2}{p^2-n^2}\right) \ , \nonumber \\
b_2^{(N)}(\tau) &=&(1-\tau)\Theta(1-\tau) \ , \\
b_4^{(N)}(\tau) &=& \left(1-\frac{\tau}{2}-\frac{\tau}{4}\sum_{p=1}
^{2 N-n} \frac{1}{N+1/2-p} \right) \Theta(2-\tau) \ , \nonumber
\end{eqnarray}
where we have rescaled the time in units of the Heisenberg time $\tau=n/t_H =
n/N$, where $t_H = 2\pi/\Delta$  and $\Delta=2\pi/N$ is the average level
spacing (for $\beta=4$ this is the nondegenerate average level spacing). The
rescaled time tends to a continuous variable in the  large $N$ limit.
($\Theta(x)$ is the step function). In a different context, related
results where obtained in Ref.\onlinecite{haake}. It is remarkable that for the
unitary ensemble the exact and asymptotic form factors coincide. For the
symplectic case, it is well known that the strong spectral correlations
produces a strong "revival" at the Heisenberg time ($\tau=1$). Asymptotically
this revival is mathematically associated to a divergence of the form factor,
since in the large-$N$ limit it takes the form $b_4 (\tau)= \lim_{N\rightarrow 
\infty} b_4^{(N)} (\tau) = [1-\tau/2+\tau\ln (|1-\tau|)/4]\Theta(2-\tau)$. 
However, this asymptotic expression does not
allow to estimate the exact magnitude of the revival, i.e. if $<|C (\tau)|^2>$
is finite at $\tau=1$ for $\beta=4$ as $N$ tends to infinity, because both the
numerator and the denominator diverge in Eq.(\ref{7}). From Eqs.(\ref{8}) we
obtain that at large but finite $N$ the average return probability for
$\beta=4$ behaves like $<|C (\tau=1)|^2> \approx \ln N/N$, i.e. it is large in
a scale $1/N$ but remains however negligible (no "macroscopic" effect). 

The second moment of the transition amplitude (i.e., the average transition
probability $<|T(\tau)|^2>$) differs from the average return probability in
that it is time independent (no transient regime). In fact, assuming that the
state $|\chi\rangle$ is statistically independent from $|\psi(0)\rangle$ we
obtain for $\beta=1$ and $\beta=2$ 
\begin{equation}\label{9}
<|T(\tau)|^2>=< |\langle\chi|\psi(\tau)\rangle|^2 > = 1/N  \ ; 
\;\;\;\;\;\;\;\;\;\;\; \beta=1,2
\ .
\end{equation} 
Analogously we find of course that the average transition probability is 
$1/(2 N)$  for $\beta=4$.

Before closing this subsection, and generalizing the discussion of 
Ref.\onlinecite{haake} concerning the powers of traces of unitary matrices,
we would like to include a random walk interpretation of the solutions of the 
time-dependent Schr\"odinger equation, Eqs.(\ref{3}-\ref{4}), and in particular 
of the average return probability, Eq.(\ref{7}).

If the phases $(\theta_1, \ldots, \theta_N)$ are considered as random 
independent variables, their multiples $n \theta_k$ are also random 
independent. Under this assumption concerning the phases, the complex numbers 
$C(n)$ and $T(n)$ in Eqs.(\ref{3}-\ref{4}) precisely define the position of a 
particle in the complex plane undergoing a random walk, the length of each of 
the $N$ steps of the walk also being a random variable with a distribution 
defined by Eq.(\ref{5}). In particular, the variance or typical square distance 
traveled by the particle after $N$ steps will be, for uncorrelated phases
\begin{equation} \label{10}
<|C_u (n)|^2> =\frac{N\delta_{n,0} + (\beta+2)/\beta}{(N+2/\beta)} \ ,
\end{equation} 
(the fact that the variance doesn't grow like $N$ as in the usual random walk 
process is related to the normalization in Eq.(\ref{5}) which implies 
$<|a_k|^2>=1/N$, i.e. an average step length which decreases with the number of 
steps. Putting $<|a_k|^2>=1$ multiplies the last equation by $N^2$, thus 
recovering the usual behaviour). In eq.(\ref{10}) we obtain a time-independent 
variance (since it doesn't depend on $n$ aside the delta dependence at the 
origin) whose value coincides with the asymptotic value of Eq.(\ref{7}).

The solutions of the time-dependent Schr\"odinger equation obtained in the 
context of RMT may also be thought of as a random walk in the complex plane, 
with the important difference that now there exist {\sl correlations} between
the angles of the different steps of the walk. The comparison of Eqs.(\ref{7}) 
and (\ref{10}) indicate that for times shorter than the Heisenberg time the 
variance of this correlated motion is smaller than the variance of 
the uncorrelated process. This is easy to understand intuitively. Indeed, it is 
well known, and this information is of course contained in the form factor, 
that there exist long range correlations between the $\theta$'s 
\cite{mehta}. For example, the two-point correlation function for $\beta=2$ 
exhibits for sufficiently large separations between phases an average repulsion
which goes like the inverse of the square of the distance. In the random walk
process, there exist therefore an average repulsion between the direction
$\theta_k$ of the k$th$ step and all the other ones. For $n=1$, the orientation
in the complex plane of one step increases the probability for the subsequent
steps to go in the {\sl opposite} direction. This mechanism of course
diminishes the variance or typical square distance traveled by the particle. 

As $n$ increases, the multiplication of the phases by $n$ and their periodicity 
contributes to progressively destroy this effect, since the repulsion may be 
transform into attraction by the multiplication and periodization process 
(depending on the relative position of the two phases in the unit circle).
At the Heisenberg time, two phases separated by a mean level spacing coincide 
when multiplied by $n=N$. Then, for times bigger or comparable to the 
Heisenberg time the product $n \theta_k$ start to behave as a random phase, 
uncorrelated with respect to the other ones, and the variance of an 
uncorrelated process should be recovered. This is exactly what is observed in 
Eq.(\ref{7}) for $\beta=1$ and $\beta=2$, since the function $b_{\beta}^{(N)} 
(n)$ monotonically reaches the value zero at $\tau \approx 1$. For $\beta=4$, 
however, the level repulsion is so strong that it produces a very rigid 
structure for short distances: the probability to observe equidistant phases
\begin{equation} \label{11}
\theta_k = 2 \pi k/N
\end{equation}
for $k=1,2,3,\ldots$ is considerable enhanced. This strong local order, which 
is practically absent for $\beta=1$ and $\beta=2$, decreases with $k$ and 
becomes irrelevant for $k=5-10$ (see for instance Fig.6 in 
Ref.\onlinecite{bohigas}). In the random walk process, all steps which 
approximately satisfied Eq.(\ref{11}) are in phase since $n \theta_k = N 
\theta_k = 2 \pi k$. This produces the observed behaviour of the variance for 
$\beta=4$ at the Heisenberg time, which grows like $\ln N/N$.

Finally, we comment on the absence of transient regime in the average 
transition probability (Eq.(\ref{9})). The difference between Eqs.(\ref{3}) and 
(\ref{4}) is that in the former $|a_k|^2$ is a real random number, while $a_k 
{\bar b}_k$ is a complex one (or real with a random sign for $\beta=1$). This 
means that the prefactor in Eq.(\ref{4}) contains a random phase on it, which 
of course suffices to destroy all the correlations between the $\theta$'s. 
The transition amplitude in quantum mechanics may then be thought of as a 
standard random walk process with uncorrelated phases. The variance obtained 
(Eq.(\ref{9})) coincides, aside from the normalization, with the usual result 
(multiplying by $N^2$ to recover the usual normalization we get a variance 
which grows linearly with the number of steps).

\subsection{Their distribution}

The distribution of the return and transition probabilities is obtained
straightforwardly from that of the amplitudes. For example, introducing the 
rescaled variable $\eta=|C (\tau)|^2 / (2 <|C (\tau)|^2 >)$, it follows 
from the Gaussian distribution of $C(\tau)$ that the distribution of the 
variable $\eta$ is given by
$$
{\cal P}(\eta) = \exp (- \eta) \ .
$$
An analogous equation is valid for the transition probability. This
distribution was in fact conjectured by Prange \cite{prange}  in his study of
the statistical properties of the form factor. As was discussed there, the
Poissonian distribution of $\eta$ implies that $|C(\tau)|^2$ and $|T(\tau)|^2$ 
are non self-averaging quantities, since the mean and the variance of $\eta$
are both equal to one and therefore the fluctuations compare to the mean do not
tend to zero in the large-$N$ limit. In order to observe this quantities, some 
additional smoothing (over time or over the parameters of the system) is
necessary. In fact, Eq.(\ref{6}) represents an interferent sum of oscillating
terms. The shortest period in the sum corresponds to an eigenphase difference
of order $\pi$, giving $T_{min} = O(1) \sim t_{erg}$ (see the concluding 
remarks for a discussion concerning $t_{erg}$). On the contrary, the longest
period is associated to differences in the eigenphases of the order of the mean
level spacing, $T_{max} \sim 2\pi/\Delta=t_H$. In order to keep the
time-dependent structure of the curve, a reasonable smoothing is over an
interval $\delta t$ such that $t_{erg}\ll\delta t\ll t_H$, thus reducing the
fluctuations by a factor of order $1/\sqrt{\delta t}$.  Averages over  the
initial state or parameters of the system may also be considered, like it was
done in the experimental studies of $<|C(\tau)|^2>$ 
\cite{leviandier,lombardi,alt}. 

\section{Concluding remarks}

In Eq.(\ref{1}), one iteration of the unitary matrix corresponds to the time it
takes to the system to explore (without filling) the available phase space.
This is because the state $U |\psi (0)\rangle$ can roughly be at  any point of
Hilbert space ($U$ is an arbitrary rotation in that space). This time,  called
the ergodic time $t_{erg}$, can be relatively large in a particular physical 
system as compared to other time scales involved in the problem, meaning that
the system is  not able to immediately explore the available phase space. For
example, in disordered  metallic systems a diffusive motion takes place before
the particle reaches the boundary of the sample. If the typical sample size is 
$L$ and the diffusion constant $D$, then $t_{erg} \propto L^2 /D$ (Thouless
time) which may be much larger than the elastic mean free time.  The  diffusive
motion is known to produce non-universal deviations from the RMT  in the  form 
factor for $t\alt t_{erg}$ \cite{altshuler}. These non-universalities can also 
manifest themselves as a system-dependent transient in Eq.(\ref{9}).
Non-universal  features will appear in $b_\beta^{(N)}(\tau)$ for short times in
general for any dynamical system \cite{berry2}, where $t_{erg}$ can roughly be
identified with the period of the shortest periodic orbit. As extensively
studied \cite{heller}, in a scale of the order of $t_{erg}$ the function
$b_\beta^{(N)}(\tau)$ may show a series of peaks at multiples of the shortest
periodic  orbits whose amplitude decreases as $\exp(-\lambda t)$, where
$\lambda$ is the classical Lyapunov exponent. For longer times, it follows from
the Gaussian distribution of the return amplitude with a variance of order
$1/N$  that statistically the probability of revivals (i.e. a return
probability of order one) in chaotic systems is exponentially small. 

Although a natural choice (relevant for experiments) of $|\psi(0)\rangle$ is a
wave packet, it is important to note that the relaxation process described here
is not restricted to such a class of initial states. Our results are general,
valid for any initial state (except of course the eigenstates). In particular,
$|\psi(0)\rangle$ could be a state completely delocalized in the basis we are
using to study the motion. 

In conclusion, we have studied the relaxation process in the time  evolution of
complex quantum systems. Aside from the short time non-universal features
already mentioned, the relaxation process described here is expected to  be
universal, applicable to a large class of physical systems including those 
whose classical limit is chaotic,  disordered metallic mesoscopic samples, high
lying  excitations of many body systems, and equivalent problems in optics or
acoustics. The  $1/N$ scaling factor of the fluctuations amplitude is a purely
normalization factor,  proportional to the typical component of the initial
state $|\psi(0)\rangle$ on the chaotic eigenstates of the system. For example,
for maps defined on a compact phase space,  $N$ is the dimension of the Hilbert
space, while for chaotic billiards $N$ is the volume  of the cavity. Rescaling
the time by the Heisenberg time we obtain a universal description, expected to
be independent of any specific feature  of the system like for example the
distribution or density of impurities or the Lyapunov exponents. In
Ref.\onlinecite{laksh} numerical results for the transition  probability for
several chaotic maps were reported. These were found to be independent of the
system considered and they all nicely agree with the present results. \\ 

We would like to thank Y. Alhassid, O. Bohigas, O. Legrand, O. Martin and N. 
Pavloff for fruitful  discussions and comments. We also thank the Institute for
Nuclear Theory at the University of Washington for its hospitality and partial
support during the completion of this work.

\end{document}